\documentstyle[12pt,epsf,epsfig]{article}
\include{graphicx}

\catcode`\@=11
\def\numberbysection{\@addtoreset{equation}{section}
\def\theequation{\thesection.\arabic{equation}}}
\numberbysection

\newcommand{\beq}{\begin{equation}}
\newcommand{\beqa}{\begin{eqnarray}}
\newcommand{\eeq}{\end{equation}}
\newcommand{\eeqa}{\end{eqnarray}}
\newcommand{\mean}[1]{\left\langle#1\right\rangle}

\def\rmd{{\rm d}}
\def\e{{\rm e}}

\def\sign{{\rm sign}}
\def\el{{\ell}}

\def\ta{\theta}

\def\fm{{f_{M}}}
\def\fz{{f_{Z}}}
\def\px{{\frac{1+x}{2}}}
\def\qx{{\frac{1-x}{2}}}

\def\gpe{{generalized persistence exponents}}

\def\wrt{{with respect to }}

\def\numberbysection{\@addtoreset{equation}{section}
\def\theequation{\thesection.\arabic{equation}}}
\numberbysection

\begin{document}
\setcounter{page}{1}
\pagestyle{plain}
\setcounter{equation}{0}
%
\ \\
\begin{center}
{\Large \bf 
Analytical results for generalized persistence \\[2mm]
properties of  smooth processes}\\[10mm]
\end{center}

\begin{center}
\normalsize
Ivan Dornic${}^{\dag,}$\footnote[1]{dornic@mpipks-dresden.mpg.de
  (Corresponding author)},
 Ana\"el Lema\^\i tre${}^{\ddag,}$\footnote[2]{lemaitre@drecam.saclay.cea.fr}, Andrea Baldassarri${}^{\ddag,\S,}$\footnote[3]{baldassa@roma1.infn.it},
 and Hugues Chat\'e${}^{\ddag,}$\footnote[4]{chate@drecam.saclay.cea.fr}\\[5mm]
{\it\dag\ Max Planck Institut f\"ur Physik Komplexer Systeme\\
N\"othnitzer Stra\ss e 38, D-01187 Dresden, Germany}\\[5mm]
{\it \ddag\ Service de Physique de l'\'Etat Condens\'e, CEA Saclay\\
 F-91191 Gif-sur-Yvette cedex, France}\\[5mm]
{\it\S\ Istituto Nazionale di Fisica Nucleare, Unit\'a di Camerino\\
Dipartimento di Matematica e Fisica, Universit\`a di Camerino\\
Via Madonna delle Carceri, I-62032 Camerino, Italy}
\end{center}
\begin{abstract}
We present a general scheme to calculate within the independent interval
approximation generalized (level-dependent) persistence properties for
 processes having a
finite density of zero-crossings. Our results are especially relevant
for the diffusion equation evolving from random initial conditions, 
one of the simplest coarsening systems. Exact results are obtained
in certain limits, and rely on a new method to deal with constrained
multiplicative processes. An excellent agreement of our analytical 
predictions with  direct numerical
simulations of the diffusion equation is found.
\end{abstract}

{P.A.C.S.: 02.50.Ey, 05.40.-a, 02.50.-r, 82.20.Mj}
\\[10mm]
\noindent  To appear in {\it J. Phys. A: Math. Gen.}
\vspace{1.cm}

\newpage

\section{Introduction}

In the past few years,
first-passage properties in nonequilibrium situations have attracted 
much interest, with numerous works  devoted to  the so-called
 persistence phenomenon (for a recent review, see
\cite{satyarevue}, and references therein).  The simplest setting might
 be within the field of phase-ordering  kinetics where, 
 as domain growth proceeds, the following question naturally arises
\cite{der1,marcos}: What is the fraction $R(t)$ of space which has always
 been  in the same phase (say the $+$ one) up to  time $t$ ? Equivalently
 for zero-temperature quenches, $R(t)$ is the probability that a given
spin has never 
flipped up to time $t$, and  is observed to decay in time as $t^{-\theta}$, 
 $\theta$ being the 
 persistence exponent. The latter can also be considered as a first-passage 
exponent, since $-\rmd R(t)/\rmd t$ represents the probability that an 
interface first sweeps over  a given point at time $t$. 
Experimental measurements of $\theta$ 
have been conducted in breath figures \cite{marcos}, and in  a system
of nematic liquid crystal akin to the $2d$ Ising model dynamics 
\cite{yurke}.
At par with the general view that thermal noise should be  asymptotically
 irrelevant in coarsening systems,
 several numerical methods \cite{derT,block1,block2,drogod98} indicate that 
 $\theta$ seems to keep  the same value for  all quenches in the
 broken-symmetry region.  Unfortunately, exact calculations  of this
 new critical exponent 
are scarce \cite{bray1,dhp} since, after integrating out at a given location
 in space the spatial degrees of freedom, one has 
to deal with an effective stochastic process
  which is generically non-Markovian. 
Perturbative techniques to calculate $\theta$ have therefore been
 developed \cite{satyaprl1,msbc96,dhz96,Oerding,perturb}, especially when a
 spin can be  associated with 
(the sign of) a Gaussian process, a situation common to 
various  closure schemes \cite{OJK82,Mazenko,brayRevue} of phase-ordering.
Other lines of investigation have been concerned with extending the notion 
of (local) persistence probability. This includes
notably the definition of a persistence exponent $\theta_{g}$ for the global
 order parameter (at $T_c$ \cite{thetaglobal},
 and below \cite{block1,block2}), the notion of block-persistence
 \cite{block1,block2} (which encompasses both $\theta$ and $\theta_{g}$), 
 and two others generalizations \cite{dorgod98,thetap},
relaxing in a different fashion the no spin-flip constraint inherent to
 the definition of $R(t)$.
 Here we shall deal with the  extension introduced in \cite{dorgod98}
(see also \cite{newman}), which is based on the 
consideration of the  local 
 mean magnetization:
\begin{equation}
M(t)={1 \over t}\int_{0}^{t}\! \rmd t' \sigma(t').
\end{equation}
This quantity is simply related to the  proportion of time spent in one
 the two possible phases by 
a given spin $\sigma \equiv \sigma_{\bf r}$. 
The distribution of $M(t)$ contains also the persistence
 probability, for
${\rm Prob}\left\{M(t)=+1\right\}=R(t)$. Yet, as soon as
$|M(t)| < 1$, changes of phase are allowed, and one observes that 
 $M(t)$ converges, when $t\to \infty$, to a broad
limiting distribution $f_{M}$,
 reminiscent of the classical arcsine law for Brownian motion,
 whose singular behaviour $f_{M}(x)\sim (1-x^2)^{\theta-1}$ 
at the edges $x = \pm 1$ of its support involves the persistence exponent
\cite{dorgod98}.
Another conceptually related route  giving access to $\theta$ is to
 consider, for a fixed value
of the level $x < 1$,
 the persistent large deviations
 $R(t,x)={\rm Prob}\left\{M(t') \ge x, \ \forall t' \le t\right\}$
 of the mean magnetization, which can be also interpreted
as a the ($+$)-persistence of the process ${\rm sign}[M(t)-x]$.
 For large times, this quantity decays as a power-law $t^{-\theta(x)}$
\cite{dorgod98}, and the original persistence exponent appears as the 
limiting case $\theta=\lim_{x \to 1^{-}}\theta(x)$ of a
continuously-varying family of \gpe.
 Let us mention that another infinite hierarchy 
of exponents  appears when one allows a spin to be ``reborn''
with a certain probability $p$ each time it flips \cite{thetap}, or
when one considers the survival properties of
(spatial) domains in coarsening systems \cite{domain}. In fact,
$R(t,x)$ displays a non-trivial structure even for the simple
(uncorrelated) binomial random walk \cite{bgl99}.

Although the existence of level-dependent exponents is by no means new
(see, e.g., \cite{slepian} for Gaussian processes, and the more recent
work \cite{perturb}),
it is physically very appealing that, as  pointed out in \cite{drogod98}, 
both the limit law $f_{M}(x)$ and the spectra $\theta(x)$  seems to be
{\it universal} for the $2d$ Ising model in the whole region $T<T_c$,  up 
 to a rescaling of $x$ by the corresponding equilibrium value of the 
magnetization, thus reflecting the fact that coarsening systems, though
perpetually out-of-equilibrium, are nevertheless in {\it local} equilibrium.

We have also  argued recently \cite{bcdl} that the global shape of the 
$\theta(x)$ curve
is a sensitive probe of the type of noise (stochastic or deterministic)
vehicled by the domain walls' motion, thereby suggesting an explanation
to the discrepancy currently reported \cite{block2} 
between the persistence exponents of the $2d$ Ising model
and of its continuous counterpart, the time-dependent Ginzburg-Landau (TDGL)
equation.

The purpose of the present work is to present a family of models for which
generalized persistence properties can be
 obtained analytically. We shall do this within the framework of the
independent interval approximation (IIA)   used  in \cite{msbc96,dhz96}
to calculate the  persistence exponent for a Gaussian random field evolving
under the diffusion equation.  The IIA has been used later to 
determine  the limit law for this system \cite{dorgod98}, and our work is a 
natural continuation of this study. Although the IIA is an uncontrolled 
approximation, limited to ``smooth'' processes
 (that is, if $\sigma(t)=\sign X(t)$, the process $\left\{X(t)\right\}_{t}$
 must be everywhere (with respect to $t$)
 differentiable),  this method has given  very accurate results, both 
for $\theta$ and $f_M(x)$. Excellent agreement will be also found
here when comparing our results with the
spectrum $\theta(x)$ obtained numerically for the diffusion equation.
In a certain limit, there will also appear an unexpected similarity
 with an exact relationship obtained for the  $\theta(x)$ exponents
of the  L\'evy-based model studied
in \cite{bbdg}, although the latter process resembles the $1d$ Ising model
 and is non-smooth in nature. 
To conclude our motivation, we mention that
some of our analytical results rely on a technique to deal with
 multiplicative random recursions (defining a so-called Kesten variable
\cite{kesten}),
 on which constraints are
enforced. Since such  variables show up in a variety of one-dimensional
 random systems \cite{derhil,callan},
it is also hoped that the method introduced here might prove helpful to
tackle new questions in this field.

The paper is organized as follows. In Section 2, we review the IIA.
Section 3 and Section 4 are respectively  concerned with the determination
 of generalized persistence properties when the number of spin-flips
is fixed (Section 3), and when the time is fixed (Section 4).
We conclude in Section 5 with a discussion.

\section{The independent interval approximation}

To fix notations, we first give a brief account of this method used in
\cite{msbc96,dhz96} to calculate the persistence exponent for
 the diffusion equation.
 Owing to dynamic scaling, 
coarsening systems with
algebraic growth laws can be rendered stationary by going over to 
``logarithmic'' time $\el=\ln{t}-\ln{t_0}$. 
The principle of the IIA is to assume that 
the  lengths of  the (logarithmic) times
 $l_{i}=\ln{t_i}-\ln{t_{i-1}}$
 between successive spin-flips $t_{1},t_{2},\ldots$ form a renewal process, 
the common probability distribution function $f(l)$
   of these intervals being determined by relating it to the
scaling form $\mean{\sigma(t_0)\sigma(t)}=a(t_0/t) \equiv A(\el)$ of the 
spin-spin autocorrelation function (which is known exactly for  the
diffusion equation, and more generally for any Gaussian process).
 The relationship reads, in  Laplace space
 ($\hat{f}(s) \equiv \int_{0}^{\infty}\! \rmd \el \e^{-s \el}f(\el)$), 
\begin{equation}
\label{loiIIA}
\hat{f}(s)={1-\mean{l}[1-s \hat{A}(s)]/2 \over 
1+\mean{l}[1-s \hat{A}(s)]/2}.
\end{equation}
 In ~(\ref{loiIIA}), the average length $\mean{l}$ between two 
zero-crossings is determined by the small-$\el$  behavior $A(\el)\approx
1-2\el/\mean{l}$ of the correlator. If $A(\el)-1 \sim \el^{\alpha}$ with 
an exponent $\alpha <1$, as
 happens in the
$1d$ Ising model (for which $\alpha=1/2$) or for some interface growth models
\cite{krugetal,kalkru} (where $\alpha$ is related to the roughness exponent),
then $A'(0)$ is infinite  and the IIA cannot be used
 as such (see, however, \cite{dharsatya}) to enumerate the spin-flips: 
any change of sign is likely to be followed
by a dense  sequence of zero-crossings (of fractal dimension $D_0=1-\alpha$
\cite{barbe}). 
 Otherwise,
 the expected large-time behavior
 $R(t)\sim t^{-\theta}\propto \e^{-\theta \el}$ of
the persistence probability  can be inferred from
  the existence of a pole at $s=-\theta$ for $\hat{f}(s)$ \cite{remark1}.  
This  procedure can be implemented in  the Ohta-Jasnow-Kawasaki closure
scheme of phase ordering \cite{OJK82} (abbreviated from now on as the
 ``diffusion equation''), where the dynamics of a spin is recast in terms of 
an auxiliary diffusing field,
\begin{equation}
\label{equadiffu}
{\partial \over \partial t}\phi({\bf r},t)=D \nabla^2 \phi({\bf r},t),
\end{equation}
evolving from zero-mean random (Gaussian, or simply short-ranged) initial
 conditions. The zeros of this diffusing field are
interpreted as representing 
the positions of  domain walls: 
$\sigma_{\bf r}(t)={\rm sign}\phi({\bf r},t)$. 
At a given location in space, the normalized process
$X(t) \equiv \phi({\bf r},t)/[\mean{\phi^2({\bf r},t)}]^{1/2}$ is still 
Gaussian,
therefore solely determined by its two-time correlation function:
\begin{equation}
c(t_{0},t)=\mean{X(t_0)X(t)}=\left[4 t_0 t \over 
(t_0+t)^2\right]^{d/4},
\end{equation}
 which
depends only on the ratio $t_{0}/t$ of the two times. The spin-spin
autocorrelation function is
$\mean{\sigma(t_0)\sigma(t)}=(2/\pi)\arcsin{c(t_{0},t)}$,
(this equality holding true for any Gaussian process),
 and reads, when expressed in terms of
the stationary timescale $\el=\ln{t}-\ln{t_{0}}$,
\begin{equation}
\label{correldiffu}
A(\el)=\frac{2}{\pi}\arcsin{\left\{[\cosh{(\el/2)}]^{-d/2}\right\}}.
\end{equation}
Plugging this back into ~(\ref{loiIIA}) gives values of the persistence
 exponent $\theta$ extremely
close compared with those obtained by a direct simulation 
of ~(\ref{equadiffu}): one finds respectively in $d=1,2,3$ space dimensions 
$\theta_{IIA}=0.1203$, $0.1862$, $0.2358$, while
 $\theta_{num}=0.1207(5)$, $0.1875(10)$, $0.2380(15)$
\cite{msbc96,dhz96}.

The IIA can also be used to calculate the limit law of the mean magnetization
\cite{dorgod98}.
  For our later purposes, it is convenient to introduce the ratios 
$z_{i}=t_{i-1}/t_{i}=\e^{-l_{i}}$ of two successive flip-times. We denote
 their
 common distribution by $f_{Z}$. The Laplace transform $\hat{f}(s)$ 
of the probability distribution  of the $l_i$'s is
nothing but the Mellin transform $\int_{0}^{1}\! \rmd z z^{s-1}f_{Z}(z)$ 
of that of the $z_{i}$'s, which implies that  
\begin{equation}
\label{fzto0}
f_{Z}(z) \approx a_0 z^{\theta-1}, \ \ z \to 0.
\end{equation} 
In ~(\ref{fzto0}), the proportionality constant $a_0$ can be computed
 from the residue at $s=-\theta$ of the Mellin-Laplace transform
~(\ref{loiIIA})).
 According to the sign of $\sigma(t)$, one can now rewrite 
 $t M(t)= \pm \left[t_1-(t_{2}-t_{1})+\ldots\right]$ as
\begin{equation}
\label{MxiW}
M(t)= \pm \left(1-2 \e^{-\xi}W_{n}\right).
\end{equation}
Here, $\xi=\ln{t}-\ln{t_{n}}$ is the length of
(logarithmic) time since the last zero-crossing, and 
 the ``weights''  $W_n$'s (which belong to the interval $(0,1)$)
are identified as Kesten variables, for  they obey
the following random multiplicative  relation:
\begin{equation}
\label{WnznWnm1}
W_{n}=1-z_{n}W_{n-1}.
\end{equation}
In the large-time limit, the random variable $W_n$ converges in distribution
to the solution of $W=1-ZW$, which means that the probability distribution
$g$ of $W$ has to satisfy the following integral equation:
\begin{equation}
\label{integralg}
g(w)=\int_{1-w}^{1}\!\rmd w' w'^{-1}g(w')\fz \left(\frac{1-w}{w'}\right).
\end{equation}
 The solution of  ~(\ref{integralg}) represents the major step for the 
determination of the limit law of the mean magnetization, since for a
 renewal process the law of the backward recurrence time $\xi$ is known:
 $\mean{\e^{\displaystyle{-s \xi}}}=(1 \! - \! \hat{f}(s))/\mean{l}s$. One of
 the very few cases where the Dyson-Schmidt integral 
equation ~(\ref{integralg}) can be inverted explicitly is when the
 distribution
$\fz$  is a pure powerlaw in $z$, that is when 
 $\fz(z)=\theta z^{\theta-1}, \ \forall z \ \in (0,1)$ \cite{callan}.
 One then finds
\begin{equation}
\label{gpure}
g(w)=B^{-1}(\theta,\theta+1)(1-w)^{\theta-1}w^{\theta},
\end{equation}
($B(a,b)=\Gamma(a)\Gamma(b)/\Gamma(a+b)$ being the usual Beta function),
 and eventually that the limit density of the mean magnetization is a Beta-law
on $(-1,1)$:
\begin{equation}
\label{loibetaM}
\fm(x)=B^{-1}(1/2,\theta)(1-x^2)^{\theta-1}.
\end{equation}
 This particular choice for $f_Z$,
proportional to the tail-behavior of the true distribution,
nevertheless contains much of the physics of the persistence phenomenon since,
in all coarsening systems, the limit density of the mean magnetization
is found to be numerically indistinguishable on its whole
 support from the law ~(\ref{loibetaM}) \cite{dorgod98,newman,drogod98,bcdl}.
 This can also be demonstrated within
the IIA for the diffusion equation (using the full law $f_Z(z)$ determined by
 ~(\ref{loiIIA})-~(\ref{correldiffu})), where one finds (by extrapolating the
large-order behavior of the moments of $M$) that the tail-behavior
$f_{M}(x) \sim (1-x^2)^{\theta-1}$ contains an overwhelming proportion
 of the full measure \cite{dorgod98}.

Let us now define the quantities of interest in the present work.
 Depending whether the number of spin-flips $n$ or the time $t$ is fixed, we
 want to calculate
\begin{eqnarray}
\label{defRnx}
R(n,x) &=&{\rm Prob}\left\{M_k \ge x, \ \forall \ k=1,2,\ldots,n\right\},
 \\
\label{defRtx}
R(t,x) &=&{\rm Prob}\left\{M(t') \ge x, \ \forall \ t' \le t\right\}.
\end{eqnarray}
In ~(\ref{defRnx}), $M_{k}=M(t_k)$ is the value of the mean magnetization at
exactly the $k$-th flip-time. These two aspects of generalized persistence
properties are different, but we shall show that they are related.

\section{Persistent large deviations of the mean magnetization when 
the number of spin-flips is fixed}

\subsection{General formalism}

Let us start with a study of  $R(n,x)$. We assume from now on that initially
$\sigma(0)=+1$, and that the the first spin-flip ratio $z_{1}$ possesses 
the same law as the others: $z_1=t_{0}/t_{1}$, with $t_0=1$. 
Since the last condition $\left\{M_n \ge x\right\}$ in ~(\ref{defRnx}) 
can only be violated ---had it been satisfied earlier--- 
when a spin is journeying in the minus phase,
that is, after an odd number of spin-flips, one has $R(2n+1,x)=R(2n,x)$.
Expressed in terms of the weights $W_{2k}$ introduced earlier, the $n$ 
conditions to be maintained at the $2n$th spin-flip time are:
\begin{equation}
\label{conditions_W2n}
R(2n,x)={\rm Prob}\left\{W_2 < \qx, W_4 < \qx, \ldots, W_{2n} < \qx \right\}.
\end{equation}
Instead of considering all the instances
taken by the
$W_{2k}$'s
 and then applying the 
constraints~(\ref{conditions_W2n}), it is sufficient to consider the ensemble
 of  constrained variables  $W_{2k}^{(x)}\in[0,\qx]$ transformed by
$W_{2k}^{(x)}=1-z_{2k}(1-z_{2k-1}W_{2k-2}^{(x)})$
(henceforth to simplify  notations we do not display the dependence of the
variables $W_{2k}$ upon $x$).
This does not define  a map on $[0,\qx]$, since some variables may fall
 outside the interval.
Let us introduce $\rho_{2k}(x)$, which represents the fraction of the
variables which are left on $[0,\qx]$ during iteration $2k$. Therefore one
 has,
\begin{equation}
R(2n,x) = \prod_{k=1}^{n} \rho_{2k}(x).
\end{equation}
The key of the argument is now to write the dynamics for the distributions 
of the constrained variables $W_{2k}$ on the interval $[0,\qx]$.
Their renormalized densities $h_{2k}$ 
($\int_{0}^{\qx}\! \rmd w  h_{2k}(w) = 1$) have now to obey
\begin{equation}
\label{rhokfini}
\rho_{2k}(x) h_{2k}(w) = \int_{1-w}^{1} \! \frac{\rmd w'}{w'}\fz
\left(\frac{1-w}{w'}\right) \int_{1-w'}^{1}\!\frac{\rmd w''}{w''}h_{2k-2}(w'')
\fz \left(\frac{1-w'}{w''}\right)
\end{equation}
The subsequent  step is to recognize that in the large-$n$ limit
the renormalized densities $h_{2n}$'s have to  reach a stationary 
distribution $h(w)$, vanishing identically for $w \ge \qx$. 
Once this stationary regime is reached, 
one has simply 
\begin{equation}
\label{def_rhox}
R(2n,x) {\sim} \left[\rho(x)\right]^{n}, \ \ n \gg 1,
\end{equation}
with $\rho(x)=\lim_{n \to \infty}\rho_{2n}(x)$  determined 
self-consistently through the solution of
\begin{equation}
\label{integralh}
\rho(x) h(w)=\int_{1-w}^{1} \! \frac{\rmd w'}{w'}\fz
 \left(\frac{1-w}{w'}\right) \int_{1-w'}^{\qx}\!\frac{\rmd w''}{w''}h(w'')
\fz \left(\frac{1-w'}{w''}\right),
\end{equation}
subject to the constraint that $h(w)$ be a normalized probability
 distribution on the interval $[0,\qx]$. 
We have to stress that  this exponential decay of $R(n,x)$ is 
in agreement with the 
expected algebraic behavior of $R(t,x)$ since, for a smooth process,
the typical number of spin-flips up to time $t$ scales  as
 $ \ln{t}/\mean{l}$. This would   not be true anymore for  the $1d$-Ising
 model (or 
for the L\'evy-based model studied in \cite{bbdg}), where the typical
number of spin-flips rather behaves  as $ t^{1/2}$ \cite{scaling_pn}
 (or $t^{\theta}$ for the L\'evy model),
and $R(n,x)$ instead goes to zero  algebraically.

\subsection{Exact solution when $\fz(z)=\theta z^{\theta-1}$}

Since the determination of the unconstrained probability density $g$ of the
$W$'s is already a difficult problem, there is no hope to solve 
(\ref{integralh}) for an arbitrary distribution $f_Z$. Nevertheless, when
 $\fz(z)=\theta z^{\theta-1}$, the structure of ~(\ref{integralh}) still
 allows for an exact calculation of $\rho(x)$, which we now present. 
 [Note that for
the probability distribution function $f(l)$ of
 intervals between spin-flips  on the
 logarithmic timescale, this choice for $f_Z(z)$ corresponds
 to a Poisson process:
$f(l)=\theta e^{-\theta l}$.]
In this case, rewriting ~(\ref{integralh}) as 
\begin{equation}
\label{integralhpure}
(1-w)^{1-\theta}h(w)={\theta^{2} \over \rho(x)}\int_{1-w}^{1}\!
 \frac{\rmd w'}{w'^{\theta}}(1-w')^{\theta-1} \int_{1-w'}^{\qx}\! 
\frac{\rmd w''}{w''^{\theta}}h(w''),
\end{equation} 
two  successive differentiations \wrt $w$ show that $h(w)$ obeys
 the following (purely local)  differential equation
\begin{equation}
\hspace{-2.5cm}
w(1-w){\rmd^2 h \over \rmd w^2}+\left[1-\theta-(3-2 \theta)w \right]{\rmd h 
\over \rmd w} -\left[(\theta-1)^2 - \theta^2/\rho(x)\right] h=0,
\end{equation}
 which is recognized as Gauss' hypergeometric differential equation.
The integral equation ~(\ref{integralhpure}) demands that $h(0)=0$, and this
 selects
 the solution 
\begin{equation}
\label{solution_h}
h(w) = {\rm Norm.} w^{\theta} {}_{2}F_{1}
\left(1- {\theta \over \sqrt{\rho(x)}},1+ 
{\theta \over \sqrt{\rho(x)}};1+\theta;w \right),
\end{equation}
where ${}_{2}F_{1}(a,b;c;w)$ is the hypergeometric series:
\begin{equation}
\label{defhyper}
{}_{2}F_{1}(a,b;c;w) \equiv
 {\Gamma(c) \over \Gamma(a)\Gamma(b)}\sum_{n=0}^{\infty}
{\Gamma(a+n)\Gamma(b+n) \over \Gamma(c+n)}{w^{n} \over n!}.
\end{equation}
[One remarks that for $x=-1$,  $\rho(-1)=1$,
and ~(\ref{solution_h}) does reduce to ~(\ref{gpure}), since
${}_{2}F_{1}(1-\theta,1+\theta;1+\theta;w)=(1-w)^{\theta-1}$.]
To determine $\rho(x)$, we need a second boundary condition, which is
 obtained by  differentiating the left-hand-side of ~(\ref{integralhpure})
 \wrt $w$, whereupon we set $w=\qx$. Then
the right-hand-side of the obtained equation 
is identically zero, since the range of the remaining integration vanishes. 
 Using the formula 
\begin{eqnarray}
\nonumber
{\rmd \over \rmd w}\left[(1-w)^{a+b-c}w^{c-1}{}_{2}F_{1}(a,b;c;w)\right]&=&
-(1-c)(1-w)^{a+b+c-1}w^{c-2} \times \\
&\times& {}_{2}F_{1}(a-1,b-1;c-1;w),
\label{derive2F1}
\end{eqnarray}
we obtain, for $\theta \neq 1$,
\begin{equation}
\label{mainresult}
{}_{2}F_{1}\left({-\theta \over \sqrt{\rho(x)}}\, ,\,
 {\theta \over \sqrt{\rho(x)}}
\, ;\, \theta \, ;\, {1- x \over 2}\right)=0.  
\end{equation}
[For $\theta=1$, Eq.~(\ref{derive2F1}) is not valid, 
but a similar identity applies, yielding
${}_{2}F_{1}(1-\rho^{-1/2}(x),1+\rho^{-1/2}(x);2;{1-x \over 2})=0$.]
Equation~(\ref{mainresult}) determines implicitly $\rho(x)$ 
as a function of $x$, for a given value of $\theta$.  
This formula can be inverted when $\theta=1/2$, to give
\begin{equation}
\label{rho12mi}
\rho(x)=
\left[\frac{2}{\pi}\arcsin{\sqrt{\qx}} \right]^2 \ \ \ (\theta=1/2).
\end{equation}
In this case, the constrained density $h(w)$ has also an explicit expression:
\begin{equation}
h(w)={\rm Norm.}\ {2 \arcsin{\sqrt{\qx}} \over \pi\sqrt{1-w}}
\sin{\left({\pi \arcsin{\sqrt{w}} \over 2 \arcsin{\sqrt{\qx}}}
\right)} \ \ \ (\theta=1/2),
\end{equation}
where the normalization factor (which depends of course on $x$) can also be
 calculated.
For other values of $\theta$, (\ref{mainresult}) is
generically multivalued, but  there is a unique  determination  of
$\rho(x)$ in the physical range $(0,1)$, and the ensuing solution
 can be obtained numerically
with an arbitrary precision. 

\begin{figure} 
\centerline{\epsfxsize=7.5 truecm
\epsfbox{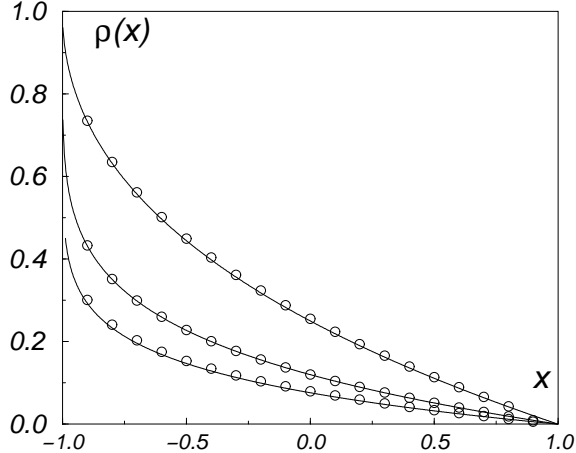}}
\caption{\label{f1}The  decay rate $\rho(x)$ of $R(2 n,x)$
(as defined by (\protect{\ref{def_rhox}})) for three representative values of 
$\theta$: $\theta=0.121$ (the value of the
persistence exponent for the $1d$ diffusion equation), 
$\theta=0.2$ (that of the $2d$ TDGL equation \protect{\cite{block2,bcdl}}),
 and $\theta=0.5$, from bottom to top. Full line: exact formula 
(\protect{\ref{mainresult}}),
circles: numerical simulations.
}
\end{figure}

Figure \ref{f1} shows our analytical prediction (\ref{mainresult}) for three
different values of $\theta$, compared with the results of a 
direct numerical
simulation of the process defined by (\ref{WnznWnm1})-(\ref{conditions_W2n}),
the $z_n$'s being distributed according to $\fz(z)=\theta z^{\theta-1}$.
The agreement found is excellent. Numerical measurements of the expected 
exponential
decay of $R(n,x)$
for the diffusion equation are hard to perform, for the
 number of spin-flips scales as $\ln{t}/\mean{l}$, which is only of the
order or $2-3$ even for times $t=10^7$.

Even if (\ref{mainresult}) is rather unwieldy for a generic value of $\theta$,
one can extract from it the behavior of $\rho(x)$ at the two edges
$x=\pm 1$ of the spectrum. This turns out to be also possible for a generic
law $\fz$, and we shall  now sketch these calculations.

\subsection{Limiting behaviour of $\rho(x)$ when $x \to  1$}

When $x\to 1$, $\rho(x) \to 0$, and  (\ref{mainresult})-(\ref{defhyper})
 show that the proper scaling variable is $S=(1-x)/2\rho(x)$. In this scaling
 limit, we have to keep all terms in the hypergeometric series 
(\ref{solution_h}), which simplifies  to
\begin{equation}
\label{Bessel}
0=\sum_{n\ge 0}\frac{\Gamma(\theta)}{\Gamma(\theta+n)}\frac{(-S)^{n}}{n!}=
\Gamma(\theta)S^{(1-\theta)/2}J_{\theta-1}(2 \sqrt{S}),
\end{equation}
where $J_{\theta-1}(v)$ is a Bessel function. The solution of (\ref{Bessel})
 is therefore
\begin{equation}
\rho(x) \approx
 \left(\frac{2 \theta}{\jmath_{\theta-1,1}}\right)^2\qx,
\end{equation}
where $\jmath_{\theta-1,1}$ is the first strictly positive zero of the Bessel
function $J_{\theta-1}(v)$. [One can check with (\ref{rho12mi}) that 
this result is correct for $\theta=1/2$, since
in this case one has simply $J_{-1/2}(v)=\cos{v}$.]
As expected, the decay rate $-\ln{\rho(x)}$ of the generalized persistence 
probability $R(n,x)$ diverges when $x \to 1$, since
no spin-flips are allowed in this limit. 

It is also possible to estimate the behavior of $\rho(x)$ when $x \to 1$ for
 the ``true'' law $\fz$ given by the inverse transform of (\ref{loiIIA}). 
 To do this, we need 
to determine the behavior of $\fz(z)$ near $z=1$. This  can be done by 
 developping (\ref{loiIIA}) for $s\to \infty$ in powers of $1/s$. The
resulting expression involves the coefficients of the Taylor series
expansion of the field-correlator near $\el=0$: $C(\el)=\mean{X(0)X(\el)}=
1+c_2\el^2/2! + c_4 \el^4/4!+\ldots$.
To lowest order in $z \to 1$,
 the result for the diffusion equation in $d$ dimensions is
\begin{equation}
\label{fzto1}
\fz(z) \approx a_1 (1-z), \ \ {\rm with} \ \
a_1=\frac{c_2^2-c_4}{2 c_2}=\frac{d+2}{32}.
\end{equation}
A  careful study of the 
integral equation (\ref{integralh}), using the fact  that the singular 
part of $g(w)$ in the limits $w \to 0,1$ is still given by (\ref{gpure})
 for any law  
 $f_{Z}(z) \approx a_0 z^{\theta-1}$, 
shows eventually that
 $\rho(x) \sim (1-x)^{2}$ when  $x \to 1$.

\subsection{Limiting behaviour of $\rho(x)$ when $x \to -1$}

In the opposite limit $x\to -1$, $\rho(x) \to 1$, and one can derive, by
 retaining the first two terms in the series (\ref{mainresult})
 (the natural expansion parameters being $1-\sqrt{\rho(x)}$ and 
$y\equiv(1+x)/2$),  the following expression:
\begin{eqnarray}
\label{rhotom1eq1}
1-\sqrt{\rho(x)}&\approx& y^{\theta}\left[{\theta \Gamma^{2}(\theta) \over
\Gamma(2 \theta)}
-y {\theta^3 \Gamma^2(\theta) \over (1-\theta)\Gamma(2 \theta)}
-y^{\theta}(1+\theta \pi \cot{\theta \pi})\right]^{-1} \\
\label{rhotom1eq2}
&\approx& \frac{1}{2}B^{-1}(\theta,\theta+1)\left(\px\right)^{\theta}
\end{eqnarray}
[The higher-order correction term in ~(\ref{rhotom1eq1}),
even though it looks superficially divergent for $\theta=1$, can be
 analytically  continued to give the correct result 
\begin{equation}
\nonumber
1-\sqrt{\rho(x)}\approx {\px \over 1-\px -\px \ln{\left(\px\right)}}, \ \
 x \to 1, \ \ \theta=1.]
\end{equation}
To lowest non-trivial order in $(1 \! + \! x)/2$, using (\ref{gpure}),
 (\ref{rhotom1eq2}) is therefore equivalent to 
\begin{equation}
\rho(x)\approx1-B^{-1}(\theta,\theta+1)\left({\px}\right)^{\theta} \approx 
\int_{0}^{\qx}\! \rmd w \ g(w), \ \ x \to -1.
\end{equation}
This result has a simple and intuitive interpretation: in this
 weakly-constrained limit, the $W_{2k}$'s obeying (\ref{conditions_W2n})
 are simply asymptotically distributed according to the {\it unconstrained}
probability distribution $g(w)$. 

Again, such a behavior is not
restricted to the particular choice  $f_Z(z)=\theta z^{\theta -1}$: for an
 arbitrary distribution behaving as in (\ref{fzto0}),
 one can show, still using the fact that
 $g(w) \sim w^{\theta}(1-w)^{\theta-1}$ when $w \to 0,1$,
that the Ansatz 
\begin{equation}
\label{rhotom1generic}
1-\rho(x) \sim \left(\px\right)^\theta, \ \ x \to -1,
\end{equation}
does solve the 
integral equation (\ref{integralh}). 

\section{Persistent large deviations of the mean magnetization when the time
 is fixed}

Even for the simplest  choice $f_Z(z)=\theta z^{\theta-1}$
(corresponding to a Poisson process with mean $1/\ta$ on the
logarithmic timescale), we have
not succeeded in calculating exactly the complete spectra $\theta(x)$
of generalized persistence exponents: when expressed in terms of
the spin-flip ratios, the constraints that one has to enforce are 
non-linear and non-local.
For an arbitrary law $\fz$, we can nevertheless offer a
 {\it systematic} perturbative development in the neighbourhood of $x=1$, and
a non-perturbative resummation in the vicinity of $x=-1$, the latter being
presumably exact (though we cannot strictly prove it).

\subsection{Limiting behavior of $\theta(x)$ when $x \to 1$}

The first idea is to follow by continuity
 $\theta(x)$ from $\theta=\theta(1)$ in 
an expansion in the number of spin-flips  which have occurred up to time
 $t$. 
 The first term of the expansion for $R(t,x)$ 
is trivial,
and just corresponds to the persistence probability. That is,
if no spin-flips have occurred up to time $t$, then the first flip-time
$t_1=1/z_1 \ge t$, and the probability of this event is $R_0=\int_{0}^{1/t}
\! {\rmd}z_{1}\fz(z_1)$. If just one spin-flip takes place up to time $t$,
then the two first flip-times satisfy  $t_1 < t \le t_2$, or equivalently
the  two first flip-ratios obey $t^{-1} < z_1 \le 1$, 
$0 \le z_2 \le (t z_1)^{-1}$.
The magnetization
$M(t)=t^{-1}(t_1-(t-t_1)]=-1+2 t_1/t$ has to satisfy $M(t)\ge x$ to
contribute to $R(t,x)$. This tightens the  upper bound on the first
flip-time ratio $z_1$, and gives a contribution 
\begin{equation}
\label{R1develop}
R_1=\int_{1/t}^{1/(yt)}\! {\rmd}z_{1}\fz(z_1)\int_{0}^{(t z_1)^{-1}}
\! {\rmd}z_2 \fz(z_2)
\end{equation}
to $R(t,x)$ (as in Section 3,  we still denote $y=(1+x)/2$; note also
that the IIA has been made by writing the joint probability distribution
of $z_1,z_2$ in factorized form).
Consider now the situation where a spin has experienced two spin-flips up
to time $t$. This corresponds to the following bounds on $z_1,z_2,z_3$:
\begin{equation}
\label{bornes3flips}
t^{-1} < z_1 \le 1, \ \ (t z_1)^{-1} < z_2 \le 1, \ \ 0 \le z_3 \le (t z_1
z_2)^{-1}.
\end{equation}
There are now two contributions to $R(t,x)$ to distinguish, according to the
range of values taken by the first flip-time ratio $z_1$. If the latter
obeys the same bounds as in (\ref{R1develop}), then, irrespectively of the
subsequent flip-time sequence $z_2,z_3$, it is not hard to check that
such an instance will always contribute
to $R(t,x)$, and the  bounds on $z_2,z_3$ are just those defined
in (\ref{bornes3flips}):
\begin{equation}
R_{2,1}=\int_{1/t}^{1/(yt)}\! {\rmd}z_{1}\fz(z_1)\int_{(t z_1)^{-1}}^{1}
\! {\rmd}z_2 \fz(z_2) \int_{0}^{(t z_1 z_2)^{-1}}
\! {\rmd}z_3 \fz(z_3) 
\end{equation}
The remaining contribution to $R(t,x)$ after two spin-flips reads
\begin{equation}
\label{R22develop}
R_{2,2}=\int_{1/(yt)}^{1}\! {\rmd}z_{1}\fz(z_1)\int_{y}^{1}
\! {\rmd}z_2 \fz(z_2) \int_{0}^{(t z_1 z_2)^{-1}}
\! {\rmd}z_3 \fz(z_3).
\end{equation}

We now evaluate   the sum of these four contributions in the long-time
limit, {\it and} for $y=(1+x)/2$ close to $1$. The derivation will show that
 we do not need to take
into account higher-order contributions in the number of spin-flips
if we content ourselves with an expansion of $\theta(x)-\theta$ to lowest
 non-trivial
order in $(1-x)/2=1-y \to 0$. 
In most of the integrals, the large-$t$ behavior is determined by
the small-$z$ form (\ref{fzto0}) of the distribution function
of the ratio of flip-times, while other spin-flip ratios can be summed over
without restriction due to the fact that $y \to 1$. For instance, to
 lowest-order, (\ref{R1develop}) simplifies to 
\begin{equation}
R_1 \approx a_0 \int_{1/t}^{1/(yt)}\! {\rmd}z_{1}{z_1}^{\theta-1}\int_{0}^{1}
\! {\rmd}z_2 \fz(z_2) \approx \frac{a_0}{\theta t^{\theta}}(y^{-\theta}-1) 
\approx \frac{a_0}{t^\theta}(1-y).
\end{equation}
This  kind of simplification is also at work for $R_0$ and $R_{2,1}$,
which are also found to be simply proportional to $t^{-\theta}$ with 
$y$-dependent amplitudes, but not for $R_{2,2}$, where the intermediate
 integral in equation (\ref{R22develop}) shows that what matters is also the
 behavior
(\ref{fzto1}) of $\fz(z)$ when $z \to 1$. One eventually  finds that
\begin{equation}
\label{R22developresult}
R_{2,2} \approx \frac{{a_0}^{2}a_1}{2  \theta}
\frac{(1-y)^2  \ln{t}}{t^{\theta}}.
\end{equation}
The appearance of a term involving a logarithm of the time is mandatory
to determine $\theta(x)$, for
if we develop
the expected algebraic behavior $R(t,x)\sim t^{-\theta(x)}$ for both
 $t \gg 1$ and $x \to 1$, we would obtain
\begin{eqnarray}
\nonumber
t^{-\theta(x)} &\approx& \exp{\left[
-\theta(1) \ln{t} +(1-x)\theta'(1)\ln{t}+\ldots\right]}\\
\label{eps_expansion}
 &\approx& t^{-\theta}\left[1-(1-x)\theta'(1)\ln{t}+\ldots\right].
\end{eqnarray}
Comparing the two expansions (\ref{R22developresult}) and 
 (\ref{eps_expansion}), one finds that for $x \to 1$
\begin{equation}
\label{taxto1}
\theta(x) \approx \theta
 \left[1-\frac{a_0 a_1}{2 \theta}\left(\qx\right)^{2}\right],
\end{equation}
where $a_0$ and $a_1$ are the coefficients describing
the behavior of $f_Z(z)$ respectively  near $z=0$ and $z=1$ (equations
(\ref{fzto0}) and (\ref{fzto1})).

\subsection{Limiting behavior of $\theta(x)$ when $x \to -1$}

The preceding subsection has shown that there is an interplay between
the distributions of the flip-time ratios, the level $x$, and the time $t$,
which appears hard to handle  in a systematic fashion. More
precisely, the distribution
of the number $N^{(x)}_{t}$ of spin-flips which have occurred up to time $t$
and which contribute to the generalized persistence probability $R(t,x)$,
depends on the level $x$. A scaling argument such as 
$\mean{N^{(x)}_{t}} \sim \mean{N^{(-1)}_{t}}=\ln{t}/\mean{l}$ --- although
qualititavely correct to explain the exponential
dependence of $R(n,x)$ and the converse
 algebraic one of $R(t,x)$ --- does not 
suffice for a quantitative calculation.
However, we expect that for $x \to -1$ we can  use the unconstrained
distribution  for the number $N_t\equiv N^{(-1)}_t$
of spin-flips which have taken
place up to time $t$, providing that we take into account the correct
form of the survival
probability $R(2n,x)$ every other time step. Namely, one should have the
 decoupling equation:
\begin{equation}
\label{handwaving}
{\rm Prob}\left\{M(t') \ge x , \ \forall t' \le t\right\} 
\approx {\rm Prob}\left\{M(t_k) \ge x, \ \forall k \le n\right\} \times
{\rm Prob}\left\{N_t=n\right\}.
\end{equation}
Taking the equation (\ref{handwaving})
for granted,
 one can express the first-passage probability at the level $x$ at time
$t$ as
\begin{equation}
\label{handwaving2}
-\frac{\rmd}{\rmd t}R(t,x) = \sum_{n=0}^{\infty} R(2n,x)
 {\rm Prob}\left\{N_t=2n+1\right\}.
\end{equation}

Using the fact that the Laplace tranform ($s \leftrightarrow \el=\ln{t}$)
of ${\rm Prob}\left\{N_{t}=n\right\}$ is equal to
 $\hat{f}^n(s)[1-\hat{f}(s))]/s$, and that the asymptotic form of the
 generalized persistence probability after a fixed number
of spin-flips is $R(2n,x) \sim
[\rho(x)]^n$,  (\ref{handwaving2}) boils down to a  geometric summation in
  Laplace space. A non-trivial pole emerges at
$s=-(\theta(x)+1)$, where 
$\theta(x)$  obeys 
\begin{equation}
\label{taxrhox}
\hat{f}(-\theta(x))={1 \over \sqrt{\rho(x)}}.
\end{equation}
It is interesting to  remark that (\ref{taxrhox})  is  exactly the equation 
 derived (also within the IIA) in \cite{thetap}  for  the exponent
$\tilde{\theta}(p)$ associated with the survival probability of a spin
which is reborn with probability $p$ each time it flips, providing that we
make the correspondence $p =\sqrt{\rho(x)}$. There is however a
 difference between 
this extension of the notion of persistence and the one we are studying here.
In our case, the survival probability $\rho(x)$ imposed 
every other timestep is not
an external parameter, but is directly inherited from the dynamics.
 Nevertheless, one expects that the difference between these two aspects
should vanish when the survival
probability (be it either $p$ or $\sqrt{\rho(x)})$ is close to $1$, thus
 constituting another plausible argument backing (\ref{handwaving})
\cite{remark2}.

Now, as $\theta(x) \to 0$ when $x \to -1$, the expansion of the 
Mellin-Laplace transform
$\hat{f}(s)\approx \hat{f}(0)+s\hat{f}'(0) \equiv 1-s \mean{l}$ for a small
argument $s=-\theta(x)$ gives
\begin{equation}
\label{tax_el_rho}
\theta(x) 
 \approx \mean{l}^{-1}\left(1-\sqrt{\rho(x)}\right)
\end{equation}
For the case $\fz(z)=\theta z^{\theta-1}$, where we have an exact expression
for $\rho(x)$, the coefficients recombine neatly 
(using in turn ~(\ref{rhotom1eq2}) and ~(\ref{loibetaM})) to give
\begin{equation}
\theta(x) \approx \frac{\theta}{2}B^{-1}(\theta,\theta+1)
\left(\px\right)^\theta \approx \theta
\int_{-1}^{x}\! {\rmd}x' f_{M}(x'),
\end{equation} 
an equality
valid again to lowest order in $x \to -1$. This seems to be the only
case where this relationship, already found in \cite{bbdg}, holds true
(thus correcting a slight imprecision made in \cite{bcdl}).
Otherwise, using (\ref{rhotom1generic}), one has simply an asymptotic
proportionality:
\begin{equation}
\label{taxtom1generic}
\theta(x) \sim (1+x)^\theta \sim \int_{-1}^{x}\! {\rmd}x' f_{M}(x'), \ \
 x \to -1.
\end{equation}
This proviso made,  we believe that, in the
limit $x \to -1$, the relationship (\ref{taxtom1generic}) between the
  spectrum of generalized persistence
exponents and the limit law of the mean magnetization  
should hold for {\it any} coarsening system, independently of the underlying
 nature
(smooth or non-smooth) of the process. Indeed, sites satisfying
 (\ref{defRtx}) must at least have their
local magnetization $M > x$, and in the long-time limit there is a fraction
$\int_{x}^{1}\!{\rmd}x' f_M(x')$ of such sites. Conversely, the decay rate
$\theta(x)$ (with respect to the logarithmic timescale $\el=\ln{t}$)
 of the generalized
 persistence probability  $R(t,x)$ should be proportional to
$\int_{-1}^{x}\!{\rmd}x' f_M(x')$,
 at least when correlations and constraints can be neglected, that is 
for $x \to -1$. The simplest
possibility (on dimensional grounds) is then that the proportionality 
factor between $\theta(x)$ and
$\int_{-1}^{x}\!{\rmd}x' f_M(x')$ be given by the usual persistence exponent
$\theta$. However, as testified by  (\ref{tax_el_rho}), other constants
 may show up when $\theta$ does not
parametrize completely the distribution of intervals between 
zero-crossings (i.e. when the latter is not exactly equal to
 $f(l)=\theta {\rm e}^{-\theta l}$, 
which implies in particular that $\mean{l}$ and
$1/\theta$  are different), which
is the generic situation for smooth processes. Note also that, for
 the L\'evy-model
studied in \cite{bbdg}, there is no other scale than the one given by $\theta$
(when the L\'evy laws have an infinite mean, i.e. for an exponent $\theta<1$),
 and --- thanks to the Sparre Andersen theorem (see, e.g., the appendix of
\cite{bgl99}) --- the equality
 $\theta(x)=\theta \int_{-1}^{x}\!{\rmd}x' f_M(x')$
holds exactly for any level $x$.

\subsection{Comparison with spectra measured for the diffusion equation}

\begin{figure} 
\centerline{\epsfxsize=7.5 truecm
\epsfbox{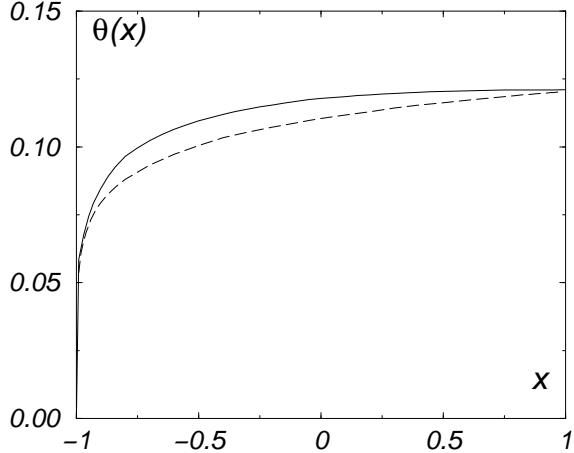}}
\caption{\label{fig2}$\theta(x)$ as measured numerically for the $1d$
 diffusion
 equation (upper curve, full line, system size $2.10^7$), and for a process
 with uncorrelated ratios of flip-times, powerlaw distributed  according to
$\fz(z)=\theta z^{\theta-1}$, with $\theta=0.121$ (lower curve, dashed line).
$500$ values for $x$ are considered in both cases.}
\end{figure}

We now turn to a comparison of our analytical results with numerical
measurements of $\theta(x)$ conducted for the  diffusion equation (Figure
\ref{fig2}), in one and two space dimensions.

The behaviors predicted by (\ref{taxto1})-(\ref{taxtom1generic}) are
 confirmed, even though they have been derived within the IIA.
At the lower edge of the spectrum, a power-law rise $\theta(x)
\sim (1+x)^{\theta}$ 
is compatible with the data (despite the obvious numerical
imprecisions in this region). At the upper edge of the spectrum, where
we have a much smaller statistical uncertainty,
$\theta(x)$  approaches its limiting value $\theta \approx 0.121(1)$ with
a vanishing slope. A numerical fit of the data of the form
  $\theta(x)-\theta \sim
(1-x)^{\mu}$ would give an exponent $\mu \approx 2.2 \pm 0.1$, which
is in fair agreement with our prediction (\ref{taxto1}).
We have also represented on Figure
\ref{fig2} the spectrum of generalized persistence exponents obtained by
direct simulation of a temporal process for which the laws of successive
spin-flip ratios are taken to be
uncorrelated and distributed according to a pure
powerlaw $\fz(z)=\theta z^{\theta-1}$, with an  exponent $\theta=0.121$. 
For such a zero-dimensional process, 
clean powerlaws extending over fifteen decades of time can be obtained
without much effort, 
and the numerical accuracy on the $\theta(x)$ spectrum is very high.
Let us note that in this case the probability density function
$\fz(z)$ does not vanish near $z=1$
 and,
browsing through the chain of
arguments leading to (\ref{taxto1}), the correct prediction near $x \to 1$
is that $\theta(x)$ reaches its limiting value with a finite slope. This is
indeed observed in our simulations.
One also remarks that, even within this ``doubly'' simplified model, the
overall agreement with the real spectra is rather good.

\begin{figure} 
\centerline{\epsfxsize=7.5 truecm
\epsfbox{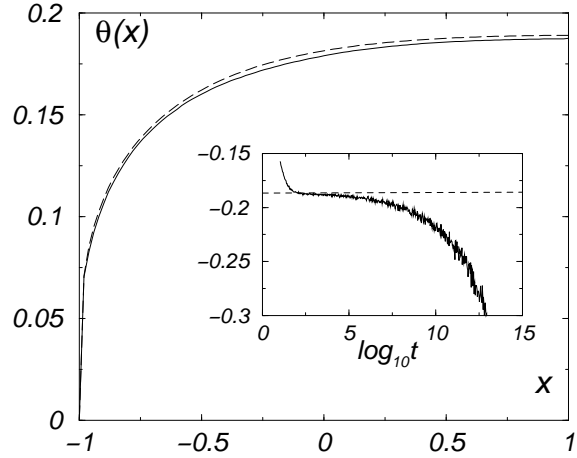}
}
\caption{\label{fig3}$\theta(x)$ as measured numerically for the $2d$
 diffusion
 equation (lower curve, full line, system size $11585^2$), and for a process
 with uncorrelated ratios of flip-times,  obtained via a rational Pad\'e
approximant of (\protect{\ref{loiIIA}}) (upper curve, dashed line).
$500$ values for $x$ are considered in both cases. The inset shows the local
slope of  the generated $f(l)$ in a log-log scale, the horizontal dashed
 line representing
the theoretical value $\theta_{\rm IIA} \approx 0.1862$.}
\end{figure}

To  quantify better the effect of temporal correlations, we have
 also determined numerically
the $\theta(x)$ spectrum for a
 process with  a distribution of
 (uncorrelated) intervals given by the full law (\ref{loiIIA}).
We have first followed \cite{dhz96}, Taylor-expanding (\ref{loiIIA})
in powers of $1/s$. Inversion term by term
gives the expansion of $f(l)$ near $l=0$, and the resulting series
 is resummed by constructing  rational Pad\'e approximants
 of the form $l P(l^2)/Q(l^2)$, $P$ and $Q$ being polynomials of respective
 degree $N$ and $N+1$. For the $2d$ diffusion equation, we have extended the
order of the Pad\'e approximants to $N=18$. 
The spin-flip  ratios are  generated using a
 rejection method, and the $\theta(x)$ can then be obtained. A comparison
with the real spectra computed for the $2d$ diffusion equation
 is presented on Figure \ref{fig3}.

On the whole range of the $x$-values, the agreement is excellent.
Of course, beyond a certain range of times, the Pad\'e approximants are not
 able anymore to mimic the exponential decay of the probability distribution
function $f(l)$ (see inset of Figure \ref{fig3}),
 which is the reason why the curve corresponding to the
 Pad\'e method is systematically overshooting the real one.

\begin{figure} 
\centerline{\epsfxsize=7.5 truecm
\epsfbox{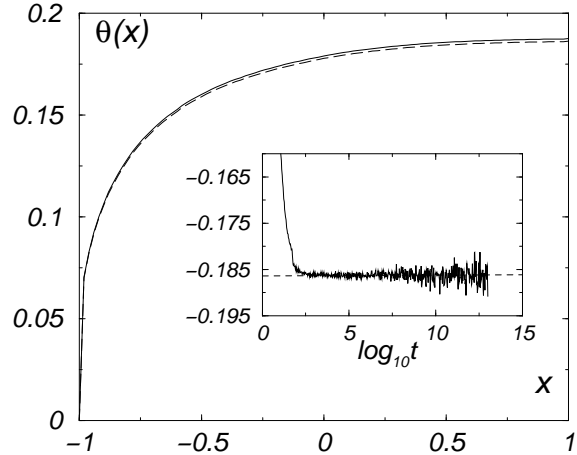}
}
\caption{\label{fig4}$\theta(x)$ as measured numerically for the $2d$
 diffusion
 equation (upper curve, full line, system size $11585^2$), and for a process
 (lower curve, dashed line) with uncorrelated ratios of flip-times,  of which
the probability distribution function
 $f_{Z}(z)=f(l=\ln{(1/z)})|{\rm d}l/{\rm d}z|$ is
 tailored to match  both the small-$l$ and large-$l$ behaviors of the law
 $f(l)$ as
encoded in (\protect{\ref{loiIIA}}). The inset shows the local
slope of  the generated $f(l)$ in a log-log scale, the horizontal dashed line
 representing
the theoretical value $\theta_{\rm IIA} \approx 0.1862$.}
\end{figure}

To circumvent this limitation,  we have eventually parametrized the
 probability distribution function $f(l)$  as 
$f(l)=(l U(l)/V(l))\exp{(-\theta l)}$, $U,V$
being polynomials in $l$ of respective order $N$ and $N+1$. The
coefficients of these polynomials are determined to match  faithfully both
 the  small-$l$ ($f(l)=a_1 l+\dots$) and large-$l$ behavior
 ($f(l) \approx a_0 e^{-\theta l}$)
of the law of the intervals. The implementation of this procedure
 (with $N=4$) for the $2d$ diffusion equation leads to a discrepancy 
(Figure \ref{fig4}) on the whole spectrum of $\theta(x)$ exponents 
which does not exceed the difference
between --- and is of the same sign as ---  the numerically measured value
  $\theta (=\theta(1)) \approx 0.1875$
and the value $\theta_{IIA}=0.1862 \dots$ predicted by the IIA.

\section{Conclusion}

We have shown in this work how it is possible to 
gain a better understanding of  persistence properties 
for processes having a finite density of zero-crossings. 
The analytical  results we have obtained, even though they  have been derived
 within an approximate scheme, are in excellent agreement with numerical
simulations conducted for the diffusion equation, which is considered as one
 of 
the simplest  --- yet non-trivial --- model of  coarsening. Furthermore, 
unexpected
connections with relationships obtained in another class of 
 stochastic processes
 have also emerged, which indicates that a certain universality
of first-passage properties may exist in those systems, at least on a
formal level.
Let us also emphasize that, if one forgets the relationship (\ref{loiIIA})
between the distribution function of intervals and the correlation function,
it becomes possible \cite{bcdl} to  account for 
all types of   $\theta(x)$ spectra encountered in coarsening systems.

Finally, outside the realm of persistence and of its siblings, the method 
we have presented in Section 3
should have a wide range of applicability for  multiplicative stochastic
 processes. For instance, the calculation  we have presented in Section 3
for generalized persistence properties after  a fixed number of spin-flips
can be rephrased in terms of the exact solution of a random fragmentation
 problem, the latter topic being of ongoing interest
\cite{derfly87, frachetal95, krapmaj00}.
It is also striking to note that first-passage exponents 
calculated recently 
for
reaction-diffusion problems in the presence of disorder with the help
of an asymptotically exact renormalization group also obey
hypergeometric equations similar to (\ref{mainresult}) \cite{monledou}.


\begin{thebibliography}{99}
\bibitem{remark1}
Actually, the persistence
probability is slightly different from  the distribution function of
 intervals, since in the former, one does not condition on having zeros at
 both endpoints of the interval. But the large-time behaviour 
($\propto \exp{(-\theta \el)}$) of these two quantities is the same.
\end{thebibliography}
\end{document}